# Apptainer Without Setuid

*Dave* Dykstra[1,1]

[1]Scientific Computing Division, Fermilab, Batavia, IL, USA

**Abstract.** Apptainer (formerly known as Singularity) since its beginning implemented many of its container features with the assistance of a setuid-root program. It still supports that mode, but as of version 1.1.0 it no longer uses setuid by default. This is feasible because it now can mount squashfs filesystems, ext3 filesystems, and overlay filesystems using unprivileged user namespaces and FUSE. It also now enables unprivileged users to build containers, even without requiring system administrators to configure /etc/subuid and /etc/subgid unlike other "rootless" container systems. As a result, all the unprivileged functions can be used nested inside of another container, even if the container runtime prevents any elevated privileges. As of version 1.2.0 Apptainer also supports completely unprivileged encryption of Singularity Image Format (SIF) container files. Performance with a particularly challenging HEP benchmark using the FUSE-based mounts both with and without encryption is essentially identical to the previous methods that required elevated privileges to use the Linux kernel-based counterparts.

## 1 Introduction

Apptainer [1] (formerly known as Singularity [2]) was designed to provide a container platform that was easy to use by scientific applications on High Performance Computing (HPC) systems. Unlike other container platforms such as Docker [3] and Podman [4], it did not attempt to support all operating system functions. These differences explain why many of its design choices were different than other container platforms. It needed to support older operating systems that didn't yet have many of the newer container-specific features of more recent Linux kernels. It needed to support containers with large numbers of small files, which when stored on an HPC's high speed filesystem can put heavy loads on the filesystem's metadata server. For those reasons and others, it was implemented using setuid-root privileges to run in namespaces and to mount filesystem image files that combined all the small files together into a larger single file. That moved the metadata operations to compute nodes, offloading the filesystem metadata server and resulting in improved performance with the container system compared to not using a container system. This was a major factor in its rapid adoption by many users.

However, using setuid-root is always a security risk, and Apptainer has supported some common use cases without setuid-root for a long time. More recently Linux kernel support for unprivileged container operations has drastically improved. Apptainer version 1.1.0 takes advantage of enough of that new kernel support that it has been made non setuid-root by default.

1 Corresponding author: dwd@fnal.gov

## 2 Non-setuid vs setuid

Apptainer still optionally supports a setuid-root mode. That can be installed by either installing the pre-built *apptainer-suid* package in addition to the *apptainer* package, or by compiling from source code with the *–with-suid* option.

Using the default non setuid-root mode has these advantages over using the setuid-root mode:

1. Setuid-root programs are notoriously difficult to fully secure. Apptainer keeps the setuid portions to a minimum and has passed a careful review, but still it is a risk.
2. Linux kernel developers believe that it is inherently unsafe to allow unprivileged users to modify an underlying filesystem while kernel code is actively accessing the filesystem [5]. Kernel filesystem drivers do not and cannot protect against all kinds of modifications to that data which it has not itself written, and that fact could potentially be used to attack the kernel. Apptainer mounts images using the kernel PR_NO_NEW_PRIVS feature which prevents the most obvious modifications for obtaining elevated privileges (by setting setuid or setcap bits), but this is a significant risk. In fact, specific vulnerabilities in the ext4 kernel module by which the kernel could be attacked have been identified [6], so Apptainer's setuid-root mode no longer allows mounting those types of filesystems by default as of version 1.1.8.
3. Non-setuid Apptainer can run nested inside another Apptainer or inside any other container runtime that disallows elevating privileges.
4. Unprivileged users can install their own copy of Apptainer without needing assistance from a system administrator.

However, there are also some disadvantages of the non setuid-root mode:

1. Mounting from unprivileged user namespaces makes use of FUSE filesystems, which run extra processes in user space. This has potentially lower performance than kernel filesystems, but performance measurements have shown that the difference for a typical HPC workflow is practically identical (see section 5 below). Metadata operations are still moved to the node running the container, so it retains that big advantage over having many small files directly on networked filesystems.
2. Supplemental groups are not fully supported with unprivileged user namespaces. The process retains the ability to write to all groups that the original user has access to, but the current group of the process cannot be changed inside the namespace.
3. Encryption of containers is done using a FUSE-based mechanism beginning in Apptainer version 1.2.0 which is not compatible with the kernel-based mechanism used previously in setuid-root mode, so encrypted containers need to be rebuilt.
4. Some little-used security and network options of Apptainer that give users elevated privileges through configuration are only available in setuid-root mode.
5. Apptainer configuration options that restrict the use of containers are not enforceable, because if unprivileged user namespaces are available then people could install their own copy and set any configuration options they choose.

Since the Linux kernel is subject to a much greater amount of scrutiny than the Apptainer setuid-root software, there have been a greater number of announced vulnerabilities that are exploitable through kernel namespace code than have been announced for Apptainer and its predecessor. Security experts generally argue that it is better to have the scrutiny than to have "security by obscurity", but urgently responding to those vulnerabilities causes additional administrator effort and can cause disruption to operations. For this reason, the Apptainer project recommends disabling network namespaces where possible, because almost all the kernel vulnerabilities over the last several years that have been exploitable via user namespaces have also required network namespaces. Apptainer does not usually use network namespaces, although other popular container systems do.

## 3 Unprivileged FUSE mounts

The primary kernel feature that non setuid-root Apptainer 1.1.0 depends on is unprivileged user namespaces. That has been available for a relatively long time, but more recently it was updated to support mounting FUSE filesystems completely unprivileged as well, even without the support of the setuid-root *fusermount* assist program. This support was even backported to Red Hat Enterprise Linux (RHEL) 7, which many HPC systems use, so it is reasonable for Apptainer to now expect that the capability is generally available. Apptainer now uses that feature to mount squashfs filesystems, ext3 filesystems, to do overlayfs when the kernel support for unprivileged overlayfs is not available, and to encrypt container image files.

### 3.1 Squashfs filesystems

Apptainer supports Singularity Image Format (SIF) container image files. SIF supports multiple partitions each with their own format, but the primary partition containing the image files is in squashfs format as supported by the Linux kernel. When running without setuid-root, Apptainer 1.1.0 mounts that partition using *squashfuse* which supports the squashfs format using FUSE. *Squashfuse* fortunately includes an option (although it was undocumented) to start at an offset into a file instead of its beginning, so it can be told where the partition starts in the SIF file.

### 3.2 Ext3 filesystems

The original Singularity container images were ext3 formatted filesystem images. Those are still supported (although not commonly used), and the format is also still supported for overlay images (which are more commonly used) both as separate files and as overlay partitions in SIF files. Apptainer 1.1.0 mounts those when running without setuid-root using *fuse2fs* which is a standard component of the *e2fsprogs* package. Unfortunately, *fuse2fs* does not include an option to start at an offset into a file, so an additional small LD_PRELOAD library was written to support mounting a partition in a SIF file.

Also unfortunately, the *e2fsprogs* package on RHEL 7 was too old to include *fuse2fs*, so it was installed as a separate *fuse2fs* package in the Extra Packages for Enterprise Linux (EPEL) yum repository for RHEL 7. It was already in a separate *fuse2fs* subpackage on all supported Debian-based systems.

### 3.3 Overlayfs

For the most common cases, Apptainer does not use any overlay system because by default image files are all in one layer and it can add bind mount points when needed by use of its lightweight (and as far as the author knows, unique) *underlay* feature. That feature makes use of scratch space for the root filesystem and bind mount points, and it bind mounts all files from the original container image that are not overridden by other bind mounts.

Apptainer images are also readonly by default, not usually allowing any other additions or modifications. It does however have an option to allow writability, and for that it uses overlayfs. The most recent Linux kernels include supporting overlayfs directly from unprivileged user namespaces, but it's not available on all operating systems that are actively used (for example it is not in RHEL7), plus overlayfs has restrictions on the types of filesystems it allows, so when overlayfs is not usable Apptainer 1.1.0 makes use of *fuse-overlayfs*.

## 3.4 Encryption

Beginning in Apptainer 1.2.0, SIF files may be encrypted with a non setuid-root installation. That makes use of *gocryptfs* which is another FUSE-based program. Unlike the kernel mechanism used for encryption in setuid-root mode, a FUSE filesystem cannot be inserted in between another filesystem and its raw bytes. Instead, *gocryptfs* expects its underlying files to be in another filesystem. Since Apptainer already supported squashfs filesystems, it was most convenient to store the underlying *gocryptfs* files in a squashfs filesystem SIF partition, even though encrypted binary data does not compress well so there's little space savings. *gocryptfs* was also found to perform best when encrypting a large file instead of many small files, so another squashfs filesystem was inserted above *gocryptfs* such that only one big file needs to be encrypted. There is a significant performance benefit to having a squashfs filesystem both above and below *gocryptfs* instead of just below, as shown in section 5, and space is also conserved because the data is compressed before encryption.

## 4 Building and running containers with fakeroot

Building containers as an unprivileged user is an especially challenging problem because that usually involves installing packages, which requires at least the appearance of running as root. Apptainer has for a long time supported building containers with the *--fakeroot* option which made use of the same "rootless" container feature used by Podman. That feature makes use of an elevated-privilege program (or two) to map user and group ids for each user based on mappings that are set up in */etc/subuid* and */etc/subgid*. However, that feature has not been enabled on most systems that use Singularity or Apptainer because it requires system administrator setup and special attention to ensure that the mappings stay consistent across clusters. Beginning in version 1.1.0, Apptainer no longer needs the mappings to enable unprivileged users to build containers.

In Apptainer 1.1.0, when an unprivileged user invokes the build command it assumes the *--fakeroot* option. It also extends the *--fakeroot* option to have additional modes that are easier to administer. Those modes are used both for building and for writable overlays. These are the ways that *--fakeroot* works depending on the conditions:

1. If the host is set up to map the current user via */etc/subuid* and */etc/subgid* mapping files, Apptainer will use that method first. This is the most complete emulation, also known as "rootless" mode, but it requires administrator setup. It also requires some elevated privilege assistance on the host, which means that it will not be able to run nested inside another container that disallows elevating privileges, as Apptainer does. Those elevated privileges on the host come from either a setuid-root install of Apptainer or via the host's *newuidmap* and *newgidmap* commands.

2. Otherwise if unprivileged user namespaces are available, Apptainer will map only the root user id to the original unprivileged user. This method is sometimes called a "root-mapped user namespace." Since this method is not as complete an emulation as "rootless mode", an info message showing it is happening is displayed.

3. If the “fakeroot” command is available on the host, Apptainer will use it in addition to a root-mapped user namespace. This command uses an LD_PRELOAD library to fake root privileges on file manipulation, telling programs that operations that would succeed as root have succeeded even though they really haven’t and keeping track of changes that would have been made by root. This is useful for avoiding errors when building containers or when adding packages to a writable container, because many package installations attempt to do additional setup that only works as root. When the fakeroot command is used, another info message is displayed showing that. The combination of a root-mapped user namespace with the fakeroot command allows most package installations to work, but the fakeroot command is bound in from the host so if the host *libc* library is a newer version than the corresponding library in the container the fakeroot command can fail. The idea to use the fakeroot command came from Charliecloud [7]. Another potential disadvantage is that this fakeroot mechanism introduces performance overhead, although containers are typically used many more times than they are built so the time to build them is not usually very important.
4. If user namespaces are not available but Apptainer has been installed with setuid-root and also the “fakeroot” command is available, then the fakeroot command will be run by itself. This allows some package installations to succeed but others will still fail; it is not as complete an emulation because the root-mapped user namespace causes the kernel to allow bypassing restrictions on files that are owned by the original user on the host, things that the fakeroot command cannot do by itself.

## 5. Performance

Measurements using a SIF-based container with a HEP benchmark called atlas-gen-bmk are shown in Figure 1. Details about the benchmark parameters and results are in [8]. The benchmark is based on python and accesses a very large number of small files which results in a very large number of metadata operations. The first measurements on a 16-core test system showed that mounting a SIF container with the unprivileged *squashfuse* was more than 6 times slower than the privileged kernel-based squashfs, an unacceptable slowdown. Another FUSE implementation in the same *squashfuse* package based on low-level FUSE called *squashfuse_ll* was not nearly as bad at only a little over 2 times slower, but it was still unacceptably bad. Very fortunately, someone had previously submitted a pull request that added multithreading support to *squashfuse_ll*, and including that patch caused the performance to be nearly the same as the kernel squashfs driver.

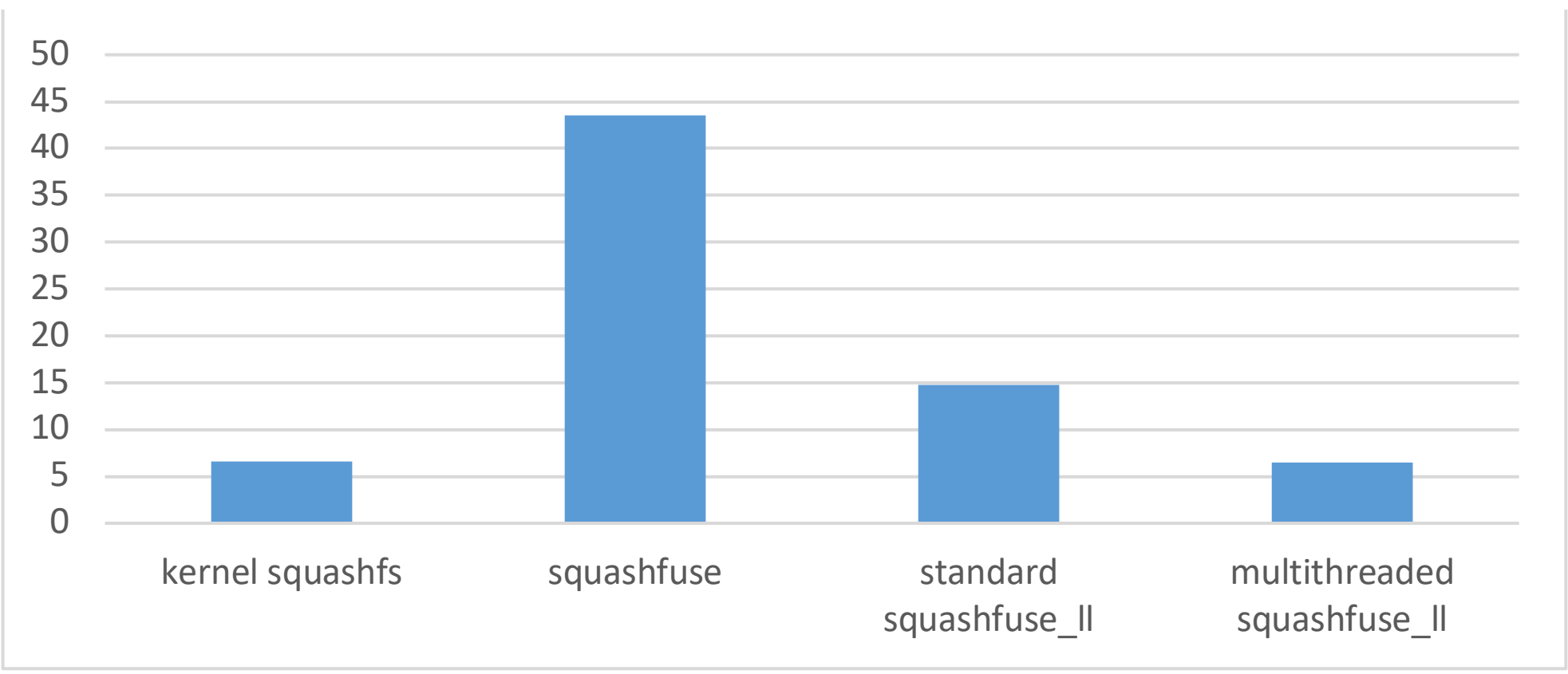

**Figure 1.** Minutes to run HEP benchmark atlas-gen-bmk with squashfs filesystem implementations

Later when the unprivileged encryption format was designed, as discussed in section 3.4 above it was found that the same benchmark performed best when it encrypted only a small number of large files instead of a large number of small files. Figure 2 shows the comparison between the benchmark run from an unencrypted squashfs filesystem, with only an additional layer of gocryptfs, and with the second squashfs filesystem on top of that, as implemented in Apptainer 1.2.0. The performance with the 3 layers is practically identical to the unencrypted performance. More details about the results are in [9].

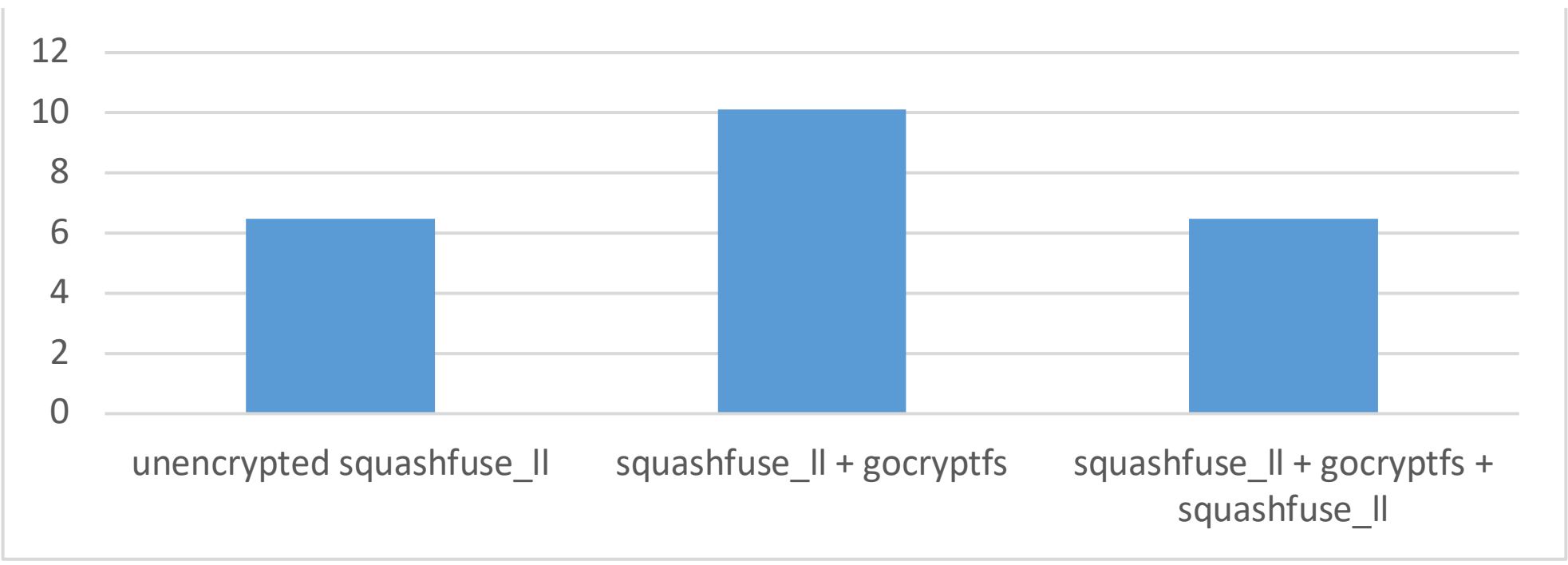

**Figure 2.** Minutes to run HEP benchmark atlas-gen-bmk with unprivileged encryption

## 6. Conclusions

From the user point of view, a default Apptainer 1.2.0 installation works with almost all of the features that they were accustomed to with previous versions, plus it enables them to build containers without needing root privileges. They can also easily install Apptainer themselves, without the assistance of a system administrator. Performance for the most common operations is practically the same as with a setuid-root installation.

From the system administrator point of view, the security risks are significantly reduced compared to a setuid-root installation, and most functionality is available without the extra effort of setting up */etc/subuid* and */etc/subgid* files.

## Acknowledgments

This builds on the work of many others, including the original author, Gregory Kurtzer. Much advice was provided by long time developers Cedric Clerget and Ian Kaneshiro. The author's work was performed using the resources of the Fermi National Accelerator Laboratory (Fermilab), a U.S. Department of Energy, Office of Science, HEP User Facility. Fermilab is managed by Fermi Research Alliance, LLC (FRA), acting under Contract No. DE-AC02-07CH11359.